\documentclass[11pt, twocolumn]{article}
\usepackage{graphicx}
\usepackage[utf8]{inputenc} 
\usepackage[english]{babel} 
\usepackage[margin=2.0cm]{geometry} 
\usepackage[titletoc,toc,title]{appendix}
\usepackage{multirow} 
\usepackage{bm} 
\usepackage[colorlinks=false]{hyperref}  
\usepackage{lscape}
\usepackage{xcolor}
\usepackage{sidecap}
\usepackage{cuted}
\usepackage{flushend}
\usepackage{hyperref}
\usepackage{gensymb}
\usepackage{mathtools, nccmath}
\usepackage{float}
\usepackage{titlesec} 
\usepackage{listings}
\usepackage{multicol}
\usepackage{lipsum}  
\usepackage{changepage}
\usepackage{float}
\usepackage{amsmath}
\usepackage{amsfonts}
\usepackage{amssymb}
\usepackage{graphics}
\usepackage{graphicx}
\usepackage{cite}
\usepackage{xcolor}
\usepackage{url}
\usepackage{hyperref}
\usepackage{pdflscape}
\usepackage{fontawesome5}
\usepackage{comment}
\usepackage{caption}
\usepackage{subcaption}
\usepackage{comment}

\definecolor{codegreen}{rgb}{0,0.6,0}
\definecolor{codegray}{rgb}{0.5,0.5,0.5}
\definecolor{codepurple}{rgb}{0.58,0,0.82}
\definecolor{backcolour}{rgb}{0.95,0.95,0.92}
\newcommand{\Like}{\mathcal{L}}

\lstdefinestyle{mystyle}{
    backgroundcolor=\color{white},   
    commentstyle=\color{codegreen},
    keywordstyle=\color{magenta},
    numberstyle=\tiny\color{codegray},
    stringstyle=\color{codepurple},
    basicstyle=\ttfamily\footnotesize,
    breakatwhitespace=false,         
    breaklines=true,                 
    captionpos=b,                    
    keepspaces=true,                 
    numbersep=5pt,                  
    showspaces=false,                
    showstringspaces=false,
    showtabs=false,                  
    tabsize=2
}

\definecolor{owngreen}{rgb}{0.0, 0.5, 0.0}

\newcommand{\changes}[1]{\textcolor{black}{#1}}

\newcommand{\github}[1]{\href{#1}{\faGithubSquare}}

\begin{document}
\title{\textbf{Cosmological parameter estimation with Genetic Algorithms}}
\author{%
  Ricardo Medel-Esquivel $^{\;1,2,3, a}$,
\and Isidro Gómez-Vargas $^{\;2, b}$  
\and Alejandro A. Morales Sánchez$^{\;4, c}$,
\and Ricardo García-Salcedo $^{\;3,5,d}$ ,
\and J. Alberto Vázquez$^{\;2, e}$
  }
  \date{%
    \small
    $^1$Escuela Superior de F\'isica y Matem\'aticas, Instituto Polit\'ecnico Nacional, 07738, Ciudad de M\'exico, M\'exico.\\
    $^2$Instituto de Ciencias F\'isicas, Universidad Nacional Aut\'onoma de M\'exico, 62210, Cuernavaca, Morelos, M\'exico.\\
    $^3$CICATA-Legaria, Instituto Polit\'ecnico Nacional, 11500, Ciudad de M\'exico,  M\'exico.\\%
    $^4$Facultad de Ciencias, Universidad Nacional Aut\'onoma de M\'exico, Ciudad de M\'exico, M\'exico.\\
    $^5$ Universidad Internacional de Valencia - VIU, 46002, Valencia, Spain. \\[1ex]%
    \large
    \today \\[1ex]
    \small
    $^a$ rmedel@ipn.mx\\
    $^b$ igomez@icf.unam.mx\\
    $^c$caracol696@ciencias.unam.mx \\
    $^d$ rigarcias@ipn.mx   \\
    $^e$javazquez@icf.unam.mx
    }


\twocolumn[
\maketitle
  \begin{@twocolumnfalse}
    \begin{abstract}  
    \changes{Genetic algorithms are a powerful tool in optimization for single and multi-modal functions. This paper provides an overview of their fundamentals with some analytical examples. In addition, we explore how they can be used as a parameter estimation tool in cosmological models to maximize the likelihood function, complementing the analysis with the traditional Markov Chain Monte Carlo methods. We analyze that genetic algorithms provide fast estimates by focusing on maximizing the likelihood function, although they cannot provide confidence regions with the same statistical meaning as Bayesian approaches. Moreover, we show that implementing sharing and niching techniques ensures an effective exploration of the parameter space, even in the presence of local optima, always helping to find the global optima. This approach is invaluable in the cosmological context, where exhaustive space exploration of parameters is essential. We use dark energy models to exemplify the use of genetic algorithms in cosmological parameter estimation, including a multimodal problem, and we also show how to use the output of a genetic algorithm to obtain derived cosmological functions. This paper concludes that genetic algorithms are a handy tool within cosmological data analysis, without replacing the traditional Bayesian methods but providing different advantages.}
    \end{abstract}
  \end{@twocolumnfalse}
]

\section{\textbf{Introduction}}
\label{sec:Introduction}
\changes{Genetic algorithms (GAs), established for decades, are tools from evolutionary computation \cite{srinivas1994genetic, tomassini1995survey, mitchell1995genetic, kumar2010genetic, katoch2021review} that solve many function optimization problems. Evolutionary computation is focused on algorithms exploiting randomness to solve search and optimization problems using operations inspired by natural evolution \cite{dumitrescu2000evolutionary}, it includes several methods for stochastic or metaheuristic optimization \cite{yang2010nature, sadeeq2023metaheuristics}; notable examples are Particle Swarm Optimization (PSO) \cite{kennedy1995particle} based on the social behavior of organisms of the same species such as birds, the Giant Trevally Optimizer (GTO) \cite{sadeeq2022giant, sadeeq2023car, hashish2023giant} inspired by the hunting behavior of predatory fish and the Artificial Rabbits Optimization (ARO) drawing inspiration from social interactions among rabbits \cite{wang2022artificial, alsaiari2023coupled}. Within evolutionary computation, the most relevant methods are genetic algorithms \cite{holland, holland1973genetic}, genetic programming \cite{langdon1998genetic}, and evolutionary strategies \cite{beyer2002evolution}; their success is due to their ability to navigate intricate, non-linear, and high-dimensional search spaces.}

\changes{In particular, Genetic Algorithms stand out as powerful tools for optimization problems because mathematically always guarantee, under certain conditions, to find the best solution, despite challenges posed by local optimum values \cite{rudolph1994convergence} and, this property puts them at an advantage over other techniques. Rooted in the emulation of natural selection and evolution, the iterative process of GAs involves generating a population, subjecting it to fitness-based selection, and applying genetic operators such as crossover and mutation. This iterative approach drives the evolution of increasingly optimal solutions over generations. GAs thrive in situations with multiple optima, irregular landscapes, or where an analytical solution is difficult to achieve. Its adaptability allows the simultaneous exploration of numerous candidate solutions, making them effective in various optimization challenges.} Unlike traditional optimization methods, GAs have the advantage of not relying on derivatives, providing excellent robustness in high-dimensional or more complex problems. Inspired by natural evolution, these algorithms efficiently explore vast and unknown search spaces \cite{garcia2020genetic}. Their ability to solve complex and dynamic projects makes them valuable in diverse fields, including medicine \cite{anastasio1998genetic, bevilacqua2001distributed, ghaheri2015applications}, epidemic dynamical systems \cite{zelenkov2023analysis, esquivel2021inverse}, geotechnics \cite{simpson1993application}, market forecasts \cite{drachal2021review}, and industry \cite{victorino2006application}, among others. A particularly successful application in the Deep Learning era is the optimization of neural networks, huge computational models in which genetic algorithms help to find optimal combinations of hyperparameters \cite{kuri2017closed, whitley1990genetic, gomez2023neural}.

With the accelerated development of computational resources, genetic algorithms, and other machine learning algorithms have been exploited in several scientific fields in recent years. Remarkably, they have resulted in significant advances in understanding particle physics \cite{abel2022evolving, bourilkov2019machine, akrami2010profile}, astronomical information \cite{charbonneau1995genetic, fridman2001radio, rajpaul2012genetic, holl2023gaia}, and cosmological phenomena \cite{axiak2011evolution, luo2020genetic, gomez2023neural_a, kamerkar2023machine, chacon2023analysis, de2022observational}.

\changes{Genetic programming, another method from evolutionary computation, has been widely used in astrophysics and cosmology \cite{Arjona_2020, nesseris2012new, wang2018computational, Bogdanos:2009ib, nesseris2010model, alestas2022machine},  which allows symbolic regression for a given data set, treating regression as a search problem to find the best combination of mathematical operators generating an expression fitting the data. Although genetic programming and genetic algorithms solve different tasks, they use similar operators to find solutions. In this work, we focus on genetic algorithms, mentioning genetic programming for reference, assuming the astrophysical community may be more familiar with it. Moreover, genetic algorithms are the most fundamental and successful evolutionary computation technique, and understanding them is useful for studying other evolutionary computation methods, including genetic programming.}

\changes{On the other hand, parameter estimation in cosmology is a very relevant task that finds a combination of values for parameters describing a cosmological model based on observational data. The goal is to refine theoretical models to align with observations for a more precise understanding of the universe. In cosmological parameter estimation, the most robust and successful algorithms are Markov Chains Monte Carlo, however, these methods sometimes are computationally expensive, and recent advancements try to attack this issue with new statistical or machine learning techniques including iterative Gaussian emulation method \cite{PellejeroIbez2019CosmologicalPE}, Adaptive importance sampling, parallelizable Bayesian algorithms \cite{Wraith2009EstimationOC}, bayesian inference accelerated with machine learning \cite{graff2012bambi, nygaard2023connect, gomez2021neural} or likelihood-free methods \cite{alsing2019fast, leclercq2018bayesian}.}

\changes{This paper aims to achieve two primary objectives: firstly, to provide a comprehensive introduction to genetic algorithms and elucidate their application in cosmological parameter estimation, and secondly, to demonstrate the complementarity of GAs with traditional Bayesian inference methods.} We include illustrative examples of optimization problems and their applications in cosmology. Particularly, we delve into using genetic algorithms to constrain the parameter space of dark energy models based on observational data. \changes{It is pertinent to mention that GAs cannot perform the same tasks as MCMC methods, and we do not try to replace them; we only perform parameter estimation with GAs by optimizing the likelihood function, whereas MCMC methods sample the posterior probability function; however, we analyze their relevance as an alternative and complementary method, as discussed in Section \ref{sec:cosmo_param_est}.}

The structure of this paper is as follows: in Section \ref{sec:biol-fund}, we present the basics of genetic algorithms and an insight into their functionality. In Section \ref{sec:applications}, we provide some examples of optimization of analytical functions by applying genetic algorithms. Section \ref{sec:cosmo_param_est} describes the path to perform cosmological parameter estimation using these algorithms. Section \ref{sec:cosmo_multi} contains examples of multimodal problems in cosmology, and in Section \ref{sec:cosmo_fgivenx}, we justify how to obtain cosmological-derived parameters from a likelihood optimization. Finally, Section \ref{sec:conclusiones} summarizes our final remarks.

\section{Fundamentals of genetic algorithms}
\label{sec:biol-fund}

\subsection{Biological fundamentals} 

Bioinspired computing is a field of computer science based on observing and imitating natural processes and phenomena to develop algorithms and computational systems \cite{bio-inspire}. These algorithms seek to solve complex problems. The bioinspired computation is classified into three main categories \cite{bio-inspire}: evolutionary algorithms (such as genetic algorithms), particle swarm intelligence (imitating collective behaviors) \cite{passino2002biomimicry, yang2010nature, mirjalili2014grey, faris2018grey}, and computational ecology (inspired by ecological phenomena) \cite{simon2008biogeography, sadeeq2023metaheuristics}. 

Genetic algorithms solve optimization \cite{srinivas1994genetic, tomassini1995survey, mitchell1995genetic, kumar2010genetic, katoch2021review} and search problems inspired by fundamental concepts of genetics and evolution \cite{mitchell1998introduction, waddington2016introduction, sadeeq2023metaheuristics}; 
some of its key points are as follows:

\begin{itemize}
    \item \textbf{Natural selection}.- Is the central principle in the theory of evolution. Just as better-adapted organisms are more likely to survive and reproduce in nature, GAs favor the fittest or most promising solutions from a population of candidate solutions. In nature, over several generations, the most promising characteristics of individuals survive to be inherited by the new generations. This is what genetic algorithms seek to do to have better solutions as more generations pass by.

    \item \textbf{Crossing}.- Also called recombination, it is a process in which genes from two parents are combined to create offspring with characteristics inherited from both parents. GAs apply the idea of crossover by combining partial solutions from two individuals in the population to generate new solutions that can inherit desirable characteristics from both parents.
  
    \item \textbf{Mutation}.- The mutation is recognized as the stochastic alterations in an organism's genetic material.  In the GAs, mutation introduces random changes in a small part of the candidate solutions, e.g., it may change the value of a bit, which increases the diversity of possible solutions and improves the exploration of the search space.
    
    \item \textbf{Reproduction and inheritance}.- In the same sense as in nature, in genetic algorithms, these operations allow the transmission of some characteristics of the parent solutions to the solutions of the next generation (offspring).
\end{itemize}

\subsection{Genetic Algorithms operations}
\label{sec:Algoritmos geneticos}

John Holland was the first to introduce the genetic algorithm in 1975. In his book \textit{Adaptation in Natural and Artificial Systems} \cite{holland, mitchell1995genetic}. According to the GA context, a population is a set of possible solutions to a given problem. Each individual has a genotype encoded in bits, then expressed as a phenotype in the problem context. The way to encode the possible solutions is fundamental to attacking a problem with GAs, and there are several options to do it, for example, with binary, integer, or real encoding, among others 
\cite{kumar2013encoding}.

Alternatively, assessing an individual's quality or a potential solution involves employing a metric or target function, ideally expected to approach its optimal value in the final generations. For the analogy of natural selection, this target function, or objective function, is called the \textit{fitness function}.  \changes{In practice, in GAs, the fitness function is directly the function to be optimized; unlike genetic programming, where the fitness function is a measure of the error between the algebraic expression found and the data set used, due to the regression task that genetic programming addresses.}

The continuous evaluation of all the individuals (possible solutions) of a population with this fitness function and the applications of genetic operations to produce new generations allow GAs to find the optimal value of this function. In the following list, we describe the fundamental procedures of genetic algorithms \cite{beasley1993overview}:

\begin{itemize}
    \item \textbf{Selection}.- It is the method of choosing the best solutions to play the role of parents and improve the quality of offspring. Several selection methods include roulette \cite{mirjalili2019genetic}, random \cite{sivanandam2008genetic}, ranking \cite{goldberg1991comparative}, tournament \cite{miller1995genetic}, and Boltzmann entropy selections \cite{lee2003entropy}.
    
    \item \textbf{Crossover}.- It is also called recombination, which generates a new possible solution given two previously selected parents. There are several crossover methods, such as one point, two points, $N$ points, uniform, three parents, random, and order. The crossover operation has an associated probability ($P_c$) that determines how many individuals recombine given the population, with $P_c=1$ indicating that all the products come from the recombination and $P_c=0$, meaning they are exact copies of the parents.

    \item \textbf{Mutation}.- After crossover, mutations make it possible to maintain diversity in the population and prevent it from stagnating at local optima \cite{marsili2000adaptive}. There are several types of crossover operators, such as flipping a gene if it is in the same position as in the parent, swapping values at random positions, flipping values from left to right, or in a random sequence and shuffling random positions. Mutation also has a probability associated with it that indicates how likely it is to randomly change a gene (bit) of a possible solution. The mutation value must be low for an efficient search within the genetic algorithm \footnote{ \changes{Let us consider a binary representation of a genetic algorithm where each individual is a sequence of binary values representing a potential solution. Suppose an individual's chromosome (binary sequence) is 101010. A mutation operation might involve flipping one of the bits, resulting in a new chromosome like 111010 or 100010. A mutation probability determines the choice of which bit to flip. If the mutation probability is low, only a few bits are expected to change, maintaining some of the original information. This process introduces diversity in the population, allowing the algorithm to explore different regions of the search space and preventing premature convergence to suboptimal solutions}.}. \changes{In the algorithm employed in our study, each bit corresponds to a specific parameter in the solution space. For instance, in the binary representation of a solution, a bit could represent the presence or absence of a particular parameter. Therefore, when we mention the likelihood of randomly changing a gene (bit) of a possible solution through mutation, we refer to the stochastic alteration of these binary digits, allowing for exploring different combinations of parameters in the search space \footnote{Consider a scenario where the objective is to determine the minimum of a straight-line model for a given set of points. The potential solutions, representing the slope ($m$) and y-intercept ($b$) of the line, are arranged with respect to the origin. If the solutions are encoded in real coding involving real numbers, the memory requirements for each input ($m$ and $b$) would depend on the bit representation of real numbers. However, by employing binary encoding, m and b can be represented as strings of zeros and ones, with each element (0 or 1) occupying only 1 bit of memory.}.}
    
    \item \textbf{Replacement}.- The last step is the replacement, which keeps the population size constant by eliminating individuals after recombination. There are three methods: strong replacement (random), weak replacement (the two fittest), and replacing both parents (the children replace both parents). 

    \item \textbf{Elitism and Hall-of-Fame}.- The elitism method ensures that the best individuals are not discarded but transferred directly to the next generation. Hall-of-Fame is an integer that indicates how many individuals are considered under elitism to be retained in the next generation. Elitism is necessary to ensure that genetic algorithms always find the best solution \cite{rudolph1994convergence}. \changes{Elitism and Hall-of-Fame are often considered distinct from the general replacement strategy. While the replacement strategy primarily focuses on selecting individuals for reproduction and forming the next generation, elitism, and hall-of-fame mechanisms specifically address preserving the best-performing individuals.}
    
    \item \textbf{Stopping criteria}.- A mechanism is needed to finalize the execution of the genetic algorithm. Some ways to do it are to stop after a fixed number of generations, after a specific time-lapse, finish the process if the best fitness does not change for several generations (steady fitness), or to stop it if there are no improvements in the objective function for several consecutive generations (generation stagnation).
\end{itemize}

In this way, we can summarize that genetic algorithms are a process that involves some crucial steps: initialization of a population form of solutions, selection of parents according to their fitness, recombination of genes by crossing, introduction of variability by mutation, substitution of individuals, and running the algorithm until the stopping criterion is satisfied. The operations described above are repeated within a loop, generation after generation until a satisfactory solution or convergence criterion is reached.

\subsection{Schema theorem}
\label{sec:Porque_funcionan?}

The heuristic search of genetic algorithms is based on Holland's schema theorem, which states that the chromosomes have patterns called schemas. This schema theorem deals with the decomposition of chromosomes into schemas and their influence on the evolutionary dynamics of the population.

A schema is a binary string of fixed length representing a chromosome pattern. For example, in a chromosome of length 6, the schema \textbf{001X00} defines a string that starts with \textbf{001}, has an unknown bit \textbf{X}, and ends with \textbf{00}.

The fitness of a schema refers to how many individuals in the population contain that specific schema. It can be represented as a fitness function $F(S)$ that denotes the fitness of the schema $S$.

The schema theorem states that high-fitness schemas are more prevalent in future generations. This is because schemas with high fitness are more likely to be selected and recombined, leading to population improvement in terms of fitness. Mathematically, we can express this as:
\begin{equation}
    F(S_{t+1}) \geq (1 - p_m) \cdot F(S_t), 
\end{equation}
where $F(S_{t+1})$ is the fitness of the schema $S$ at the next generation $(t+1)$, $F(S_t)$ is the fitness of the schema $S$ in the current generation $(t)$, and finally, $p_m$ is the mutation probability.

This equation indicates that the fitness of the schema in the next generation is at least equal to the current fitness, modulated by the mutation probability. If $p_m$ is low, schemas with high fitness will likely survive and propagate in future generations, contributing to population improvement.

\begin{table}[h]
    \captionsetup{justification=raggedright, singlelinecheck=false, font=footnotesize}
    \resizebox{8cm}{!} {
     \begin{tabular}{l} 
     \hline
     \textbf{Algorithm 1. Simple Genetic Algorithm}\\ [0.4ex] 
      \hline
      Parents $\leftarrow$ \{randomly generated population\}\\
      \textbf{While} not (termination)\\
      \; \; \; \; Calculate the fitness of each parent in the\\ 
      \; \; \; \; population\\
      \; \; \; \; Children $\leftarrow \emptyset$\\ 
      \; \; \; \; \textbf{while} |Children| $<$ |Parents|\\
      \; \; \; \; \; \; \;Use fitness to probabilistically select a pair of\\ \; \; \; \; \; \; \;parents for mating\\
      \; \; \; \; \; \; \;Mate the parents to create children $c_1$ and $c_2$\\
      \; \; \; \; \; \; \;Children $\leftarrow$ Children $\cup \{c_1, c_2\}$\\
      \; \; \; \; \textbf{Loop}\\
      \; \; \; \; Randomly mutate some of the children\\
      \; \; \; \; Parents $\leftarrow$ Children\\
      Next generation\\
      \hline     
    \end{tabular}
    }
    \caption{Pseudocode of a genetic algorithm.}
    \label{pseudocode_table}
 \end{table}

\section{Genetic algorithms application}
\label{sec:applications}

In this section, we implement a genetic algorithm to optimize univariate functions and extend its application to higher-dimensional problems. The general structure of a genetic algorithm is provided in the pseudocode of Table \ref{pseudocode_table}.
\\

Several libraries incorporate genetic algorithms, such as Distributed Evolutionary Algorithms (\texttt{DEAP}) \cite{DEAP}, \texttt{Karoo GP} \cite{Karoo_GP}, Tiny Genetic Programming \cite{Tiny_Genetic_Programming}, and Symbiotic Bid-Based GP \cite{Symbiotic_Bid_Based_GP}. These libraries simplify the implementation of genetic algorithms. In this paper, we have utilized the \texttt{DEAP} library, which boasts comprehensive documentation.

\subsection{Single variable functions}

Considering the following three functions: 
\begin{itemize}
    \item $f_1(x) = (x^2 + x) \cos(2x) + x^2$,
    \item $f_2(x) = \sin^2(3x + 45) + 0.9 \sin^3(9x) - \sin(15x + 50) \cos(2x - 30)$, 
    \item $f_3(x) = -x^6/60 - x^5/50 + x^4/2 + 2x^3/3 - 3.2x^2 - 6.4x$,
\end{itemize}
we aim to use a custom genetic algorithm to find their global maxima. 

\begin{figure*}[h]
\centering
    \captionsetup{justification=raggedright, singlelinecheck=false, font=footnotesize}
    \includegraphics[scale=0.32]{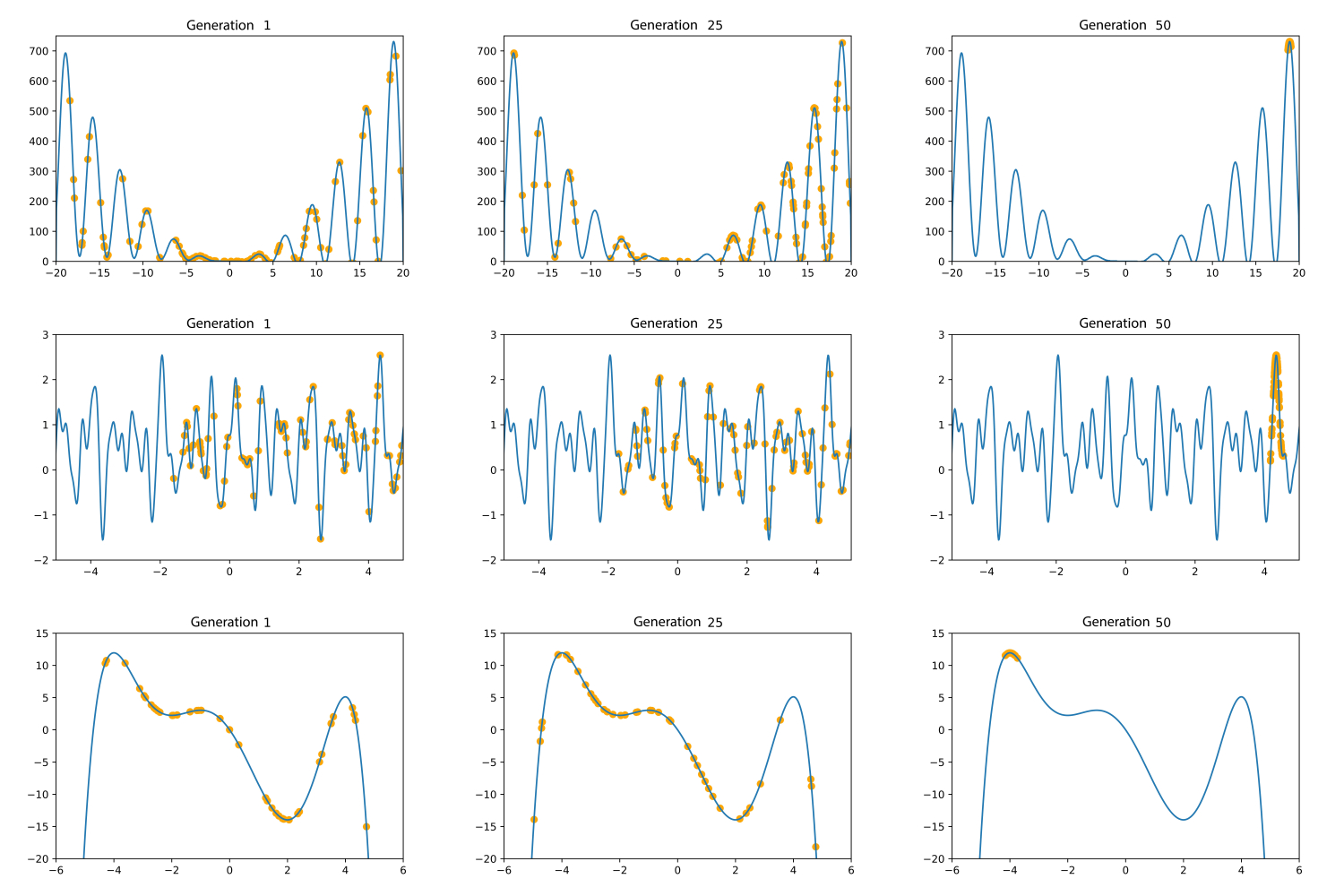}
    \caption{The search space exploration is presented for three different generations: 1, 25, and 50. As we advance through the generations, a greater concentration of individuals is seen at the global maxima. In the top panels, $f_1(x)$.  In the central panels, $f_2(x)$ In the bottom panels: $f_3(x)$}
    \label{fig:f3_poblacion}
\end{figure*}

\changes{In Figure \ref{fig:f3_poblacion}, it can be seen how the above functions are optimized by a genetic algorithm, using a population size of 100 individuals, with Hall-of-fame size equal to 1, mutation probability of 0.2 and crossover probability of 0.5, over 50 generations. Note that as the generations progress, the individuals are closer to the global maxima. Another interesting feature is that, despite local optima, the genetic algorithm in all functions can find global optima, as it is mentioned in the Introduction section and the Ref. \cite{rudolph1994convergence}.}

\label{sec:algoritmo_cosmologia}
\subsection{\textbf{Multi-modal functions}}
\label{sec:multi}

Genetic algorithms can also address problems with multiple dimensions and maxima by modifying the representation of candidate solutions and the operators used to generate new solutions. They can explore complex search spaces efficiently and identify global or local optima by appropriately designing crossover and mutation operators and analyzing different encoding techniques.

We use the Himmelblau function to demonstrate how genetic algorithms can be used to optimize these types of multi-modal functions. We use the \texttt{DEAP} library, a robust Python framework for evolutionary computation, to achieve our goal. The following equation defines the Himmelblau's function:
\begin{equation}
    f(x,y) = (x^{2} + y - 11)^{2} + (x+y^{2}-7)^{2}.
\end{equation}
The niching and sharing technique is employed to identify all global optima within a single genetic algorithm run. This concept draws inspiration from nature, where regions are divided into sub-environments or niches, enhancing population efficiency and survival. Individuals compete for resources in these niches independently of those in other niches. By integrating a sharing mechanism into the genetic algorithm, individuals are incentivized to explore new niches, discovering multiple optimal solutions, each considered as a niche. Typically, this is achieved by dividing an individual's fitness value by the sum of distances from all other individuals. This approach penalizes overpopulated niches by distributing the local rewards among their individuals \cite{wirsansky2020hands}.

\changes{Niching involves dividing the population into subpopulations, each assigned to explore a specific region in the solution space. This encourages diversity by allowing genetically engineered individuals to compete for fitness locally. Conversely, sharing ensures a fair distribution of fitness resources among individuals within the same niche. An individual's fitness is influenced not only by its performance but also by the performance of its neighbors, preventing overemphasis on a specific region and promoting a balanced exploration. This approach prevents premature convergence to a local maximum, allowing simultaneous exploration of different regions and ultimately facilitating the identification of the global maximum.}
Applying this technique effectively requires a larger population size and more generations than a simple genetic algorithm. This is essential to spread the population across the sample space, targeting different niches and, consequently, identifying multiple optimal maxima. In our experiment, we executed the algorithm with 200 individuals and 200 generations, and the outcomes are summarized in Table \ref{tab:results_niching}.

\begin{figure*}[h]
\captionsetup{justification=raggedright, singlelinecheck=false, font=footnotesize}
\centering
\makebox[12cm][c]{
    \includegraphics[trim= 0mm 0mm 0mm 0mm, clip, width=6cm]{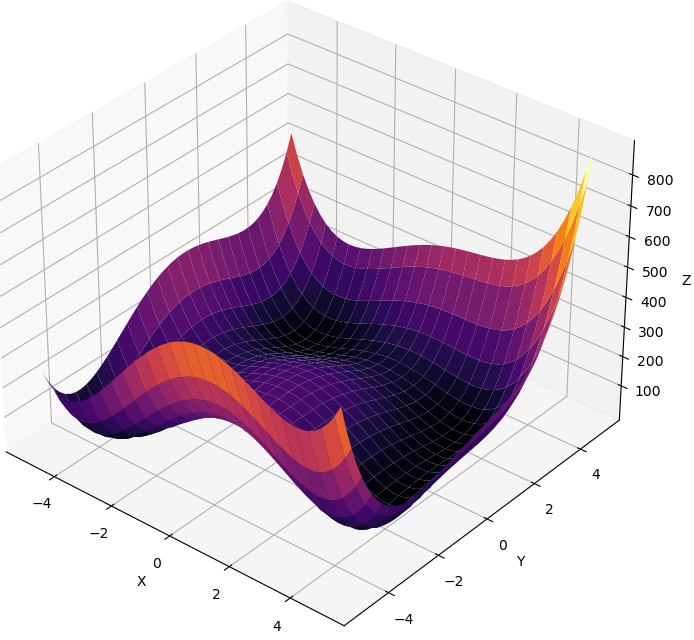}
    \includegraphics[trim= 0mm 0mm 0mm 0mm, clip, width=5cm, height=6cm]{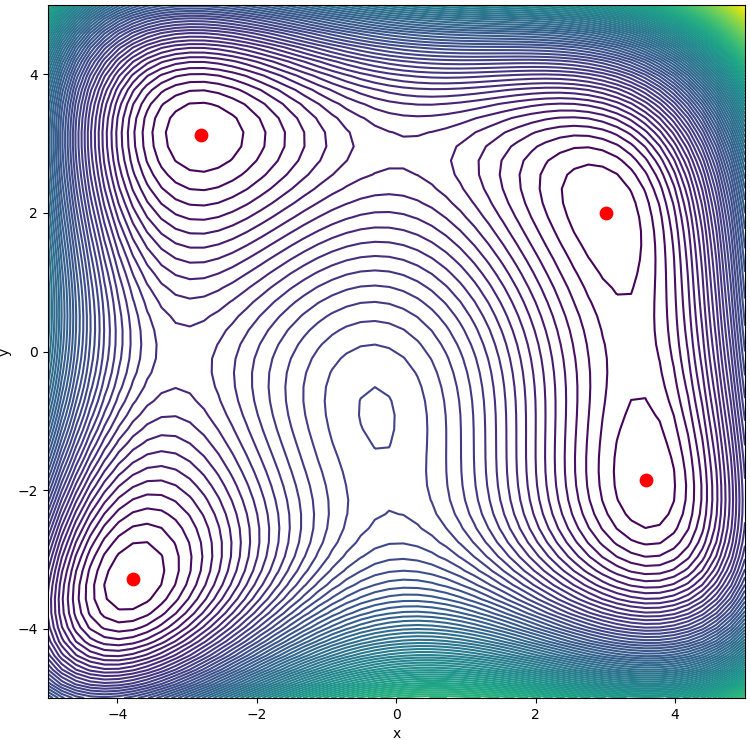}
    \includegraphics[trim= 0mm 00mm 0mm 0mm, clip, width=5cm, height=6cm]{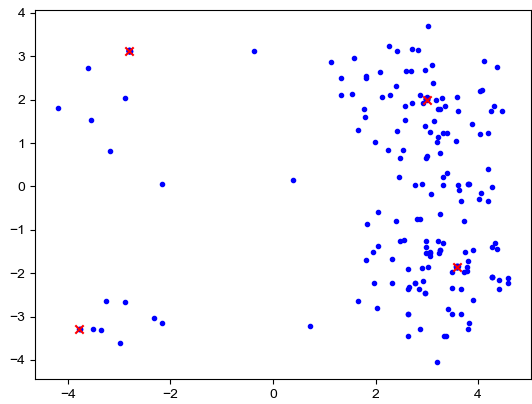}
}
\caption{On the left panel, we have Himmelblau's function, while the center panel displays its contour diagram. The red points on the contours represent the global minima of the function. On the right panel, we can observe the application of the genetic algorithm with niching and sharing, specifically for Himmelblau's function.}
\label{fig:himmel}
\end{figure*}

\begin{table}[h!]
\captionsetup{justification=raggedright, singlelinecheck=false, font=footnotesize}
    \centering
    \begin{tabular}{|c|c|}
    \hline
         Real optimum & Optimum found by GA  \\
    \hline
    \hline
         $(3.000, 2.000)$ & $(3.010, 1.998)$ \\
    \hline
         $(-2.805, 3.131)$ &$(-2.802, 3.133)$ \\
    \hline
         $(-3.779, -3.283)$ & $(-3.774, -3.292)$\\
    \hline
         $(3.584, -1.848)$ & $(3.585, -1.847)$\\
    \hline
    \end{tabular}
    \caption{A comparison is made among the four real global optima of Himmelblau's function \cite{wirsansky2020hands} and those found by the genetic algorithm using niching and sharing.}
    \label{tab:results_niching}
\end{table}

As can be seen in Table \ref{tab:results_niching}, these results are remarkably similar to the real values. Improving these results is possible by increasing the number of individuals and generations. It should also be noted that this technique is not limited to three dimensions but can be generalized to $N$ dimensions and support the search for global $M$ optima. However, it is important to remember that as the number of dimensions increases, more computational resources are required to search effectively.

\label{sec:algoritmo_cosmologia}
\subsection{\textbf{Statistical Analysis}}
\label{sec:sta}

\changes{Genetic algorithms are handy tools in statistical applications for optimizing likelihood functions, thereby determining the parameters of a scientific model (which is precisely what this article aims to demonstrate). However, reporting a confidence interval for the output of a genetic algorithm can be more complex than in classical statistical methods. The most rigorous technique relies on having a mathematical model of the genetic algorithm's convergence that extends beyond Holland's schema theory for the simple genetic algorithm published in 1975.}

\changes{Because the state of the population in a genetic algorithm depends solely on the previous state in a probabilistic manner, Markov chains have been studied as suitable models for specific applications, and more recently, others have been modeled as martingales \cite{eiben1999theory, oliveto2014runtime}.}

\changes{However, it is possible to resort to less rigorous techniques. One approach is to assume a distribution for the optimized parameters. For instance, assuming the parameters follow a normal distribution, the confidence interval can be calculated based on standard deviations, and confidence ellipses can be computed using Fisher matrices. This is the procedure employed in this article. Another procedure involves using the Bootstrap method or other re-sampling techniques \cite{vsileny1998earthquake}.}

\section{\textbf{Application in cosmology}}
\label{sec:algoritmo_cosmologia}

In observational cosmology, one of the fundamental tasks is to determine the values of the free parameters for a given theoretical model based on observational measurements. This involves creating a function that captures discrepancies between observed data and theoretical predictions and using it to obtain a parameter estimate that fits the data well. The likelihood function is typically used to represent the data's conditional probability given the theory and its parameters. Although Bayesian inference is the most robust method for parameter estimation in cosmology, as it allows sampling the posterior probability of parameters given the data, it can be computationally intensive (see nomenclature under the Bayesian formalism of the Bayes' theorem \cite{esquivel2021introduction, hogg2018data}); instead of sampling the posterior probability function to estimate parameter values efficiently, optimization algorithms can be used to find the maximum likelihood function. In Reference \cite{hogg2018data}, there is an exciting overview of the difference between 
sampling and optimization, and it can be seen that they are two different tasks that can be complementary. This section presents three applications that show how genetic algorithms can be applied to analyze cosmological data. First, we offer parameter estimation in three cosmological models: $\Lambda$CDM, CPL, and PolyCDM. We then discuss how genetic algorithms can be used in a cosmological model with multiple maximum values, such as the Graduated Dark Energy Model presented in Ref. \cite{akarsu2020graduated}.

The datasets utilized in this section comprise 31 cosmic chronometers (CC) \cite{jimenez2003constraints, simon2005constraints, stern2010cosmic, moresco2012new, 
zhang2014four, moresco2015raising, moresco20166, ratsimbazafy2017age}, Baryon Acoustic Oscillation measurements (BAO) \cite{alam2017clustering, ata2018clustering, blomqvist2019baryon, de2019baryon, beutler20116df, ross2015clustering}, 1048 Type Ia supernovae (SNeIa) sourced from the Pantheon compilation \cite{scolnic2018complete}, and binned data from the Joint Light Analysis SNeIa compilation \cite{betoule2014improved}.
\\

Considering the datasets mentioned above, we employ the following log-likelihood functions for Bayesian inference and optimization methods:

\begin{equation}
\log \Like_i = -\frac{1}{2}(D^i_{\mathrm{th}}-D^i_{\mathrm{obs}})^T \cdot C_i^{-1} \cdot (D^i_{\mathrm{th}}-D^i_{\mathrm{obs}}),
\end{equation}
where the index $i$ ranges from 1 to 3, corresponding to the three datasets: cosmic chronometers [$D^{i=1} = H(z)$], BAO [$D^{i=2} = D_A(z)$], where $D_A(z)$ represents the Hubble, volume averaged and angular distance, and SNeIa [$D^{i=3} = \mu(z)$], where $\mu(z)$ denotes the distance modulus. In this context, $D_{\rm obs}$ represents the observed measurements, while $D_{\mathrm{th}}$ represents the theoretical values for the cosmological models. The matrices $C_{i}$ encompass the covariance information, accounting for systematic and statistical 
errors.


We implemented a module to work with the \texttt{DEAP} genetic algorithms within the \texttt{SimpleMC} \footnote{https://igomezv.github.io/SimpleMC} code for our cosmological parameter estimation \cite{SimpleMC}. In some of the subsequent results, we compare the genetic algorithm's outcomes with those of Bayesian inference obtained using the nested sampling algorithms, a specialized type of Markov Chain Monte Carlo (MCMC) technique\cite{esquivel2021introduction, skilling2004nested}. Additionally, we utilize the Fisher matrix formalism described in Refs. \cite{padilla2021cosmological, sivia2006data} to calculate the confidence intervals and generate error plots for the genetic algorithm-based parameter estimation. It is important to emphasize that genetic algorithms are not employed to generate posterior samples; instead, they are used to explore maximum likelihood estimation, which can yield similar and quicker results than parameter estimation. However, they cannot replace the robustness of MCMC methods. Furthermore, we conducted maximum likelihood estimation using a classical optimization method, specifically the L-BFGS algorithm \cite{zhu1997algorithm}, for comparison purposes and to assess the advantages of genetic algorithms.

\subsection{Cosmological Parameter estimation}
\label{sec:cosmo_param_est}

As previously mentioned, we employ genetic algorithms to evaluate their effectiveness in parameter estimation. As a proof of the concept, and for simplicity, we consider three cosmological models: $\Lambda$CDM, CPL, and PolyCDM,  which are described below:

\begin{itemize}
    \item \textbf{$\Lambda$CDM}. The $\Lambda$CDM model serves as the standard cosmological model and comprises two primary components: Cold Dark Matter (CDM), which plays a pivotal role in the universe's structure formation, and dark energy, which exhibits a counter-gravitational behavior, leading to the universe's accelerated expansion. The cosmological constant, denoted by $\Lambda$, is the simplest and most straightforward representation of dark energy, which exerts a pressure equal in magnitude but opposite in sign to the universe's energy density ($p=-\rho$). For a flat universe in the late stages of its evolution, the equation governing its expansion is given by $H^2\equiv \left(\frac{\dot a}{a}\right)^2 = \rho_{m}(t)+ \rho_{\Lambda}(t)$, where 
    $a$ represents the scale factor, the dot denotes the derivative with respect to time, $\rho_{m}$ signifies the density of dark matter and baryons and $\rho_{\Lambda}$ accounts for the dark energy content in the form of a cosmological constant. These two parameters describe the evolution of the universe's content. Incorporating their initial conditions denoted with a subscript 0, this equation can be re-expressed in terms of the redshift $1+z= 1/a$ as follows:
\begin{equation}
    H^2=  H_0^2[ \Omega_{\rm CDM,0}(1+z)^3+ \Omega_{\Lambda,0}],
    \label{eq:Friedman}
\end{equation}
where $H_{0}$ denotes the Hubble constant, providing the present rate of expansion of the Universe. The parameters $\Omega_{\rm CDM,0}$ and $\Omega_{\Lambda,0}$ are specific to the $\Lambda$CDM model. The former represents the current dimensionless density of dark matter (plus baryons), while the 
latter signifies the dimensionless density of dark energy. These parameters are subject to the constraint $\Omega_{\rm CDM,0}+\Omega_{\Lambda,0}=1$; when this equality holds, we have a flat universe {\cite{liddle2015introduction}}. Consequently, for this model, we effectively have two free parameters, namely, $h$ and $\Omega_{\rm CDM, 0}$, which we simplify by denoting $\Omega_{\rm CDM}$ as $\Omega_m$ for brevity.

\item \textbf{CPL model}. One can discern dark energy's characteristics by investigating its state equation, denoted as $w(z)$, where $p$ and $\rho$ represent the pressure and dark energy density, respectively \cite{linden2008test}. Chevallier, Polarski, and Linder introduced the following parameterization for the equation of state: $w(z)=w_{0}+w_{a}\frac{z}{1+z}$, where $w_{0}$ signifies the current value of the equation of state. In contrast, $w_{a}$ represents its rate of change over time \cite{linden2008test}. This equation of state leads to the following derivation:
\begin{equation}
\begin{aligned}  
    H(z)^2 & = H_0^2 [\Omega_{m,0}(1+z)^3 + \\ 
     & (1-\Omega_{m,0})(1+z)^{3(1+w_0+w_a)} e^{-\frac{3w_a z}{1+z}}].
\end{aligned}
\label{eq:hzcpl}
\end{equation} 

Now, the parameter estimation consists of finding the free parameters $H_{0}$, 
$\Omega_{m,0}$ and $w_{0}$ and $w_{a}$.

\item \textbf{PolyCDM}. We can consider an extension of dynamical dark energy by introducing spatial curvature, $\Omega_1$, which adapts to the evolution of dark energy at low redshifts \cite{ANN_Cosmologial}. By performing a Taylor series expansion of the equation \ref{eq:Friedman} \cite{Vazquez:2012ag}, we arrive at the PolyCDM model:

\begin{equation}
\begin{aligned}
H^{2} ={} &  H_{0}^{2}(\Omega_{m,0}(1+z)^{3} + \\& \Omega_{1,0}(1+z)^{2} + \Omega_{2,0}(1+z) \\
& +(1-\Omega_{m,0}-\Omega_{1,0}-\Omega_{2,0})),
\end{aligned}
\label{eq:PolyCDM}
\end{equation}
where $\Omega_{m,0}$ represents the dark matter, and baryon, contribution and $\Omega_{2,0}$ can be interpreted as the ''lost matter'' \cite{Vazquez:2012ag}. PolyCDM can be considered a parametrization of the Hubble parameter { \cite{Zhai_2017}}.
 
\end{itemize}

\begin{figure*}[t!] 
\centering
    \captionsetup{justification=raggedright,singlelinecheck=false,font=footnotesize}
    \makebox[11cm][c]{
            \includegraphics[trim=0mm 0mm 0.0mm 0mm, clip, width=6.2cm, height=6cm]{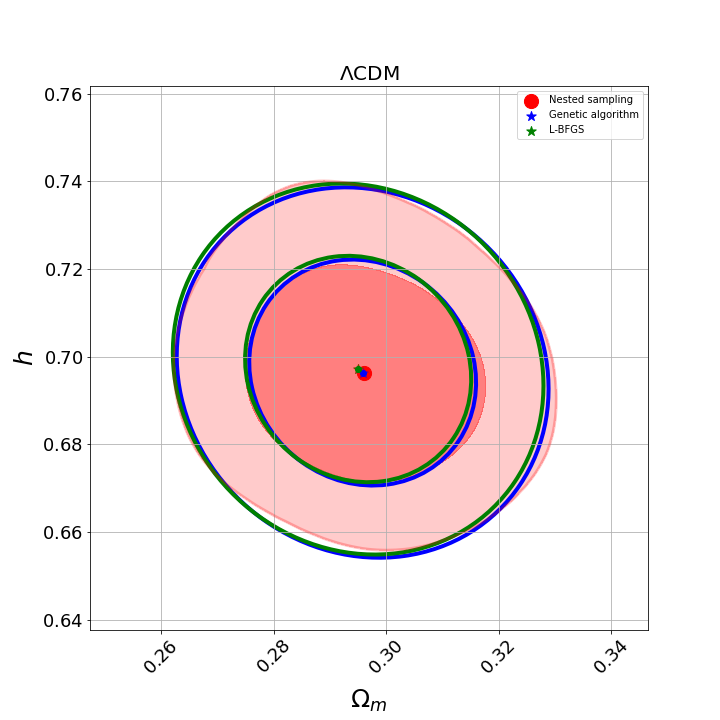}
            \includegraphics[trim=0mm 0mm 0.0mm 0mm, clip, width=6.2cm, height=6cm]{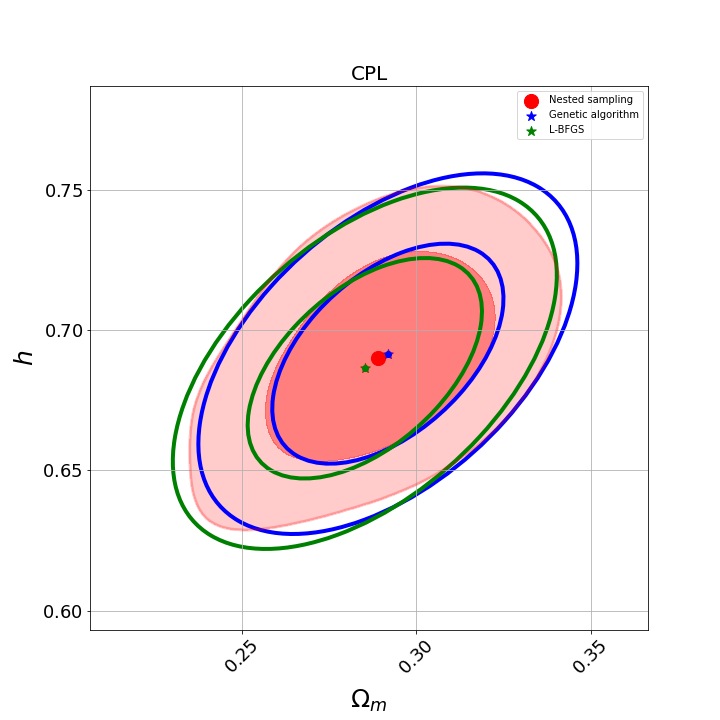}          
            }
    \makebox[11cm][c]{
            \includegraphics[trim=0mm 0mm 0.0mm 0mm, clip, width=6.2cm, height=6cm]{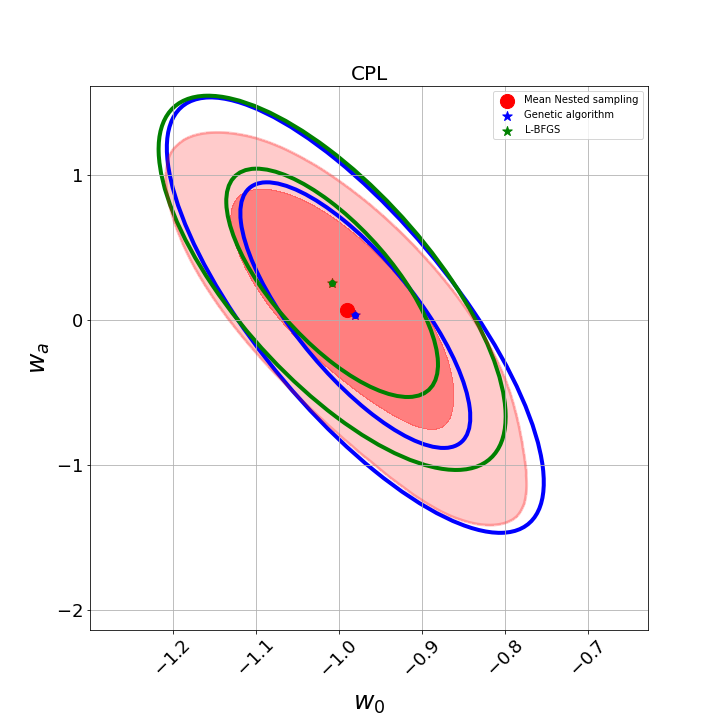}
            \includegraphics[trim=0mm 0mm 0.0mm 0mm, clip, width=6.2cm, height=6cm]{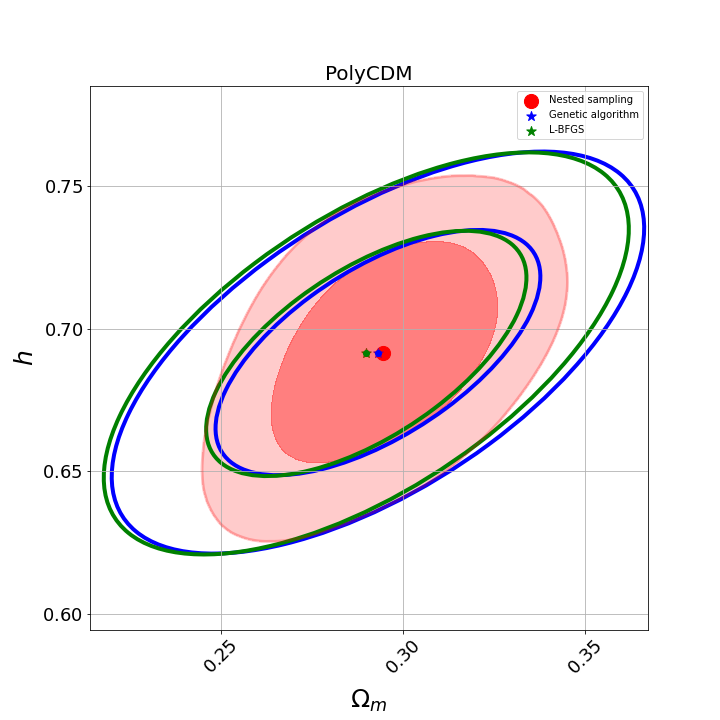}
            \includegraphics[trim=0mm 0mm 0.0mm 0mm, clip, width=6.2cm, height=6cm]{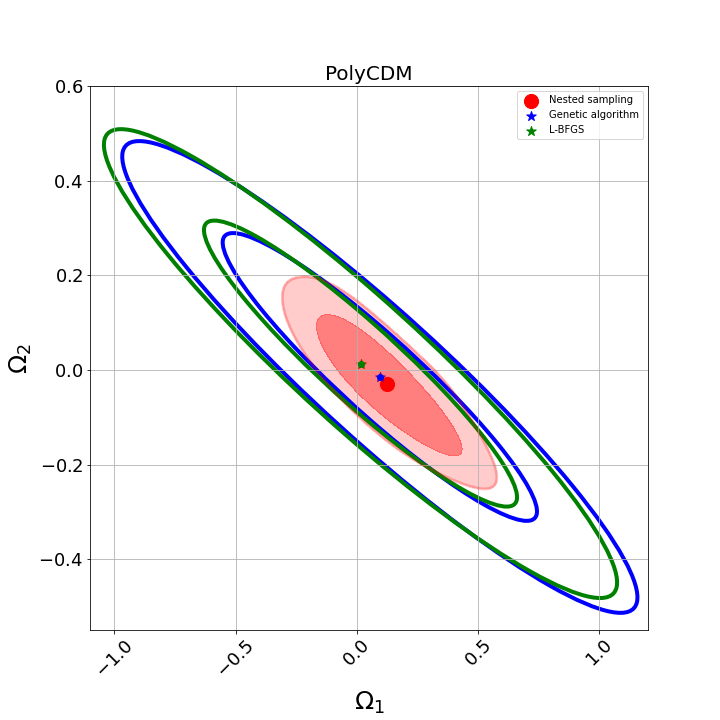}
            }
   
        \caption{2D posterior distribution plots showing the parameter mean estimates 
        from nested sampling and the parameter values obtained through likelihood 
        maximization using the L-BFGS and genetic algorithm methods (see color labels). \changes{Note that the confidence intervals are different due to their nature: optimization methods that maximize the likelihood function (L-BFGS and genetic algorithms) make use of the Fisher matrix formalism to approximate the errors (see Section \ref{sec:sta}), while the MCMC (nested sampling) method constructs its confidence intervals from sampling the posterior probability function. }}
        \label{fig:posteriorall}
    \end{figure*}

For all the models mentioned above, we use a genetic algorithm with elitism, using 50 generations, a mutation probability of 0.2, a crossover probability of 0.7, a population comprising 100 individuals, and a Hall-of-Fame size of 2 to maximize the likelihood probability function. Table \ref{tab:simple_mc_results} and Figure \ref{fig:posteriorall} present the parameter estimation results obtained throughout the three methods outlined earlier. 
It is noticeable that, in most cases, the genetic algorithm results closely align with the parameter estimations derived from the nested sampling. Consequently, although they are slower than optimization methods like the L-BFGS method, genetic algorithms offer greater precision while remaining faster than MCMC algorithms. It is important to note that genetic algorithms maximize the likelihood function rather than sampling the posterior distribution. This distinction can be computationally advantageous compared to Bayesian inference procedures in specific scenarios. However, GAs lack the assignment of weights to individuals, as found in Bayesian inference samples, and their exploration of parameter space differs from MCMC methods, which rely on Markov Chains and probabilistic conditions. Genetic algorithms, instead, focus on achieving improved solutions in each generation.

\begin{table*}[t!]
\captionsetup{justification=raggedright, singlelinecheck=false, font=footnotesize}
\centering
\scriptsize
\begin{tabular}{|c|c|c|c|c|}
\hline
\multicolumn{1}{|l|}{}   &               & \multicolumn{3}{c|}{Data: CC+BAO+SNeIa}                                   \\ \hline
Model                  & Parameters   & L-BFGS optimizer  &  Genetic             & Nested \\ \hline
\multirow{4}{*}{$\Lambda$CDM}     
                & $h_{0}$ & $0.6972 \pm 0.0170$ & $0.6964 \pm 0.0170$ & $0.6963 \pm 0.0160$ \\ \cline{2-5} 
                    
                & $\Omega_{m}$ & $0.2950 \pm 0.0133$ & $0.2958 \pm 0.0133$ & $0.2960 \pm 0.0134$ \\ \cline{2-5} 
                
                & $-2\log \Like$ & $1049.2424$ & $1049.2476$ & $1049.2445$
                \\ \hline
                
                \multirow{5}{*}{CPL}     
                & $h_{0}$ & $0.6864 \pm 0.0259$ & $0.6916 \pm 0.0258$ &  $0.6901 \pm 0.0240$ \\ \cline{2-5} 
                & $\Omega_{m}$ & $0.2853 \pm 0.0221$ & $0.2919 \pm 0.0218$ & $0.2892 \pm 0.0211$ \\ \cline{2-5} 
                & $w_0$ & $-1.0082 \pm 0.0840$ & $-0.9803 \pm 0.0912$ &  $-0.9909 \pm 0.0861$ \\ \cline{2-5} 
                & $w_a$ & $0.2556 \pm 0.5188$ & $0.0330 \pm 0.6035$ & $0.0679 \pm 0.5296$ \\ \cline{2-5} 
                & $-2\log \Like$ & $10483.9018$ & $1049.0778$ & $1048.9415$
                \\ \hline
                
                \multirow{5}{*}{PolyCDM} 
                & $h_{0}$ & $0.6913 \pm 0.0283$ & $0.6916 \pm 0.0283$ & $0.6916 \pm 0.0250$\\ \cline{2-5} 
                & $\Omega_{m}$ & $0.2899 \pm 0.0290$ & $0.2931 \pm 0.0294$ & $0.2945 \pm 0.0198$ \\ \cline{2-5}
                & $\Omega_{1,0}$ & $0.0150 \pm 0.4254$ & $0.0947 \pm 0.4271$ & $0.1232 \pm 0.1795$ \\ \cline{2-5} 
                & $\Omega_{2,0}$ & $0.0136 \pm 0.1995$ & $-0.0147 \pm 0.2007$ & $-0.0298 \pm 0.0903$ \\ \cline{2-5} 
                & $\Omega_k$ & $-0.0013 \pm 0.0703$ & $-0.0076 \pm 0.0702$ & $-0.0004 \pm 0.0117$ \\ \cline{2-5} 
                & $-2\log \Like$ & $1049.0688$ & $1049.0660$ & $1049.1286$
                \\ \hline

\end{tabular}
\caption{Parameter estimation via genetic algorithms for the $\Lambda$CDM, CPL, and PolyCDM models utilizing cosmic chronometers, BAO, and SNeIa datasets. The $-2\log \Like$ value represents the optimal fitness value.}
\label{tab:simple_mc_results}
\end{table*}

\subsection{Multimodal models}
\label{sec:cosmo_multi}

Parameter inference in some models can lead to the identification of multiple optima, meaning that posterior probability functions can have multimodal distributions. To address this complexity, Bayesian nested inference algorithms, such as Multinest \cite{feroz2009multinest}, are a sampling method designed to deal with multimodal distributions, allowing effective sampling of the parameter space. In contrast, classical optimization algorithms are limited to finding a single maximum. Genetic algorithms, thanks to niche and sharing techniques (see Section \ref{sec:multi}), have the ability to exhaustively explore the parameter space, even in the presence of local maxima. An example of a model with multiple maxima in its posterior distribution is the case of Graduated 
Dark Energy \cite{akarsu2020graduated}, which is governed by the following Friedmann equation:
\begin{equation}
\begin{aligned}
    H^2 = {} & H_0^2[ \Omega_{r,0}(1+z)^{-4}+  \Omega_{m,0}(1+z)^{-3} + \\ 
    & \Omega_{\rm DE,0} sgn [1 - \psi \ln a]|1 - \psi \ln a|^{\frac{1}{1-\lambda}}],
\end{aligned}
\end{equation}
where $\Omega_{\rm DE,0}$ is the dimensionless density parameter of the Dark Energy with $\psi < 0$ and $\lambda = 0, -2, -4, ...$. $\psi$ is defined in terms of $\lambda$ and another parameter $\gamma$ in the following way: $\psi \equiv -3\gamma (\lambda -1)$.  One maximum value corresponds to the $\Lambda$CDM model, whereas the other is present to alleviate the Hubble tension. This model resembles a rapid transition of the Universe from anti-de Sitter vacua to de Sitter vacua; see the details of the model in the references \cite{akarsu2020graduated, Acquaviva:2021jov, Akarsu:2021fol, Akarsu:2023mfb, akarsu2023relaxing}.

\begin{figure}[t!]
    \captionsetup{justification=raggedright, singlelinecheck=false, font=footnotesize}
    \centering
    \includegraphics[trim = -2mm  0mm 0mm 0mm, clip, width=4.2cm, height=4.2cm]{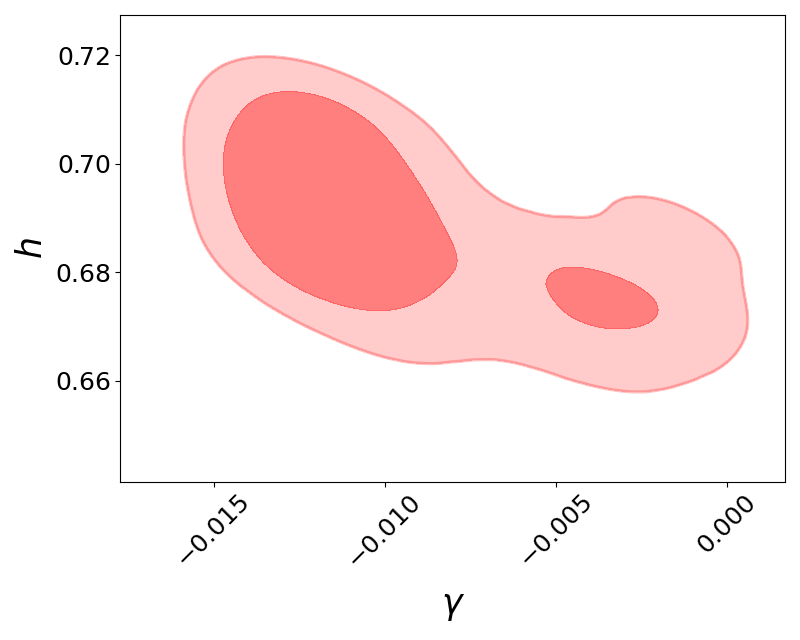}
    \includegraphics[trim = -2mm  0mm 0mm 0mm, clip, width=4.2cm, height=4.2cm]{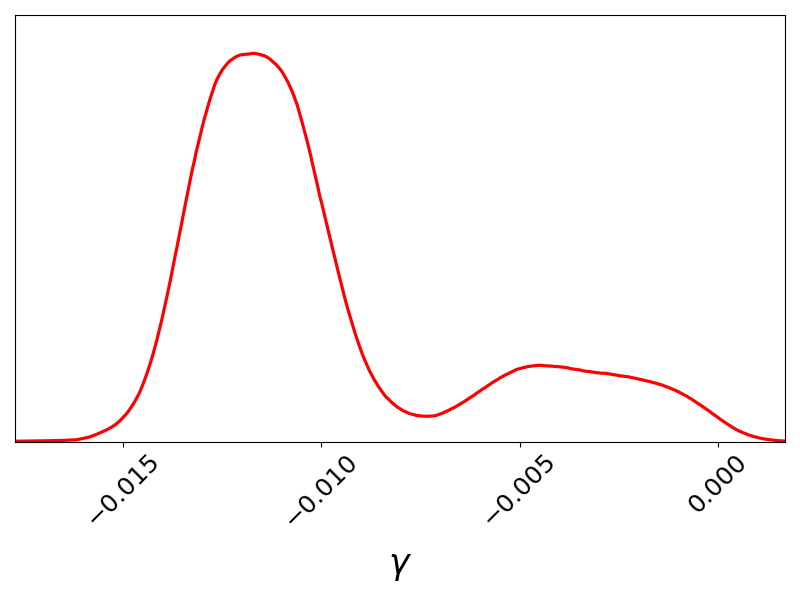}
    \caption{Posterior plots with nested sampling for $h$ and $\gamma$ parameters of the Graduated DE model using HD+BAO+SN, where the bi-modality is shown. \textit{Left:} 2D posterior plot for $h$ vs $\gamma$. \textit{Right:} 1D posterior distribution plot for $\gamma$ parameter.}
    \label{fig:graduated_posterior_nested}
\end{figure}

For the genetic algorithm with elitism used in this case, we set 20 generations, 200 individuals for the population, crossover and mutation probabilities of 0.5 and 0.2, respectively, and a Hall-of-Fame of size 2. Therefore, the free parameters for the graduated Dark Energy model are $\Omega_{m,0}$, $h_0$, $\lambda$, and $\gamma$. For this example, to appreciate the multimodality in the graduated DE model, we use the same data that in the original work (Ref. \cite{akarsu2020graduated}), i.e., cosmic chronometers, BAO and  SNeIa (binned data from the Joint Light Analysis compilation \cite{betoule2014improved}), but for simplicity, we do not use the Planck information. We also fix $\lambda=-20$.  Performing Bayesian inference to this model, the posterior distribution for $\gamma$ parameter is shown in Figure \ref{fig:graduated_posterior_nested}, in which two modes exist. In Table \ref{tab:graduated}, we can analyze the outputs of the parameter estimation using nested sampling through a posterior distribution sampling, the L-BFGS optimization method, and a genetic algorithm maximizing the likelihood distribution function; we can notice that the results maximizing the likelihoods are roughly consistent with the parameter estimation with Bayesian inference, however, for the $\gamma$ value the L-BFGS method is unable to find a value different of zero and it is far from the estimation of this parameter using the same data.

\begin{figure}[t!]
    \captionsetup{justification=raggedright, singlelinecheck=false, font=footnotesize}
    \centering
    \includegraphics[trim= -2mm 0mm 0mm 0mm, clip, width=9cm, height=6.5cm]{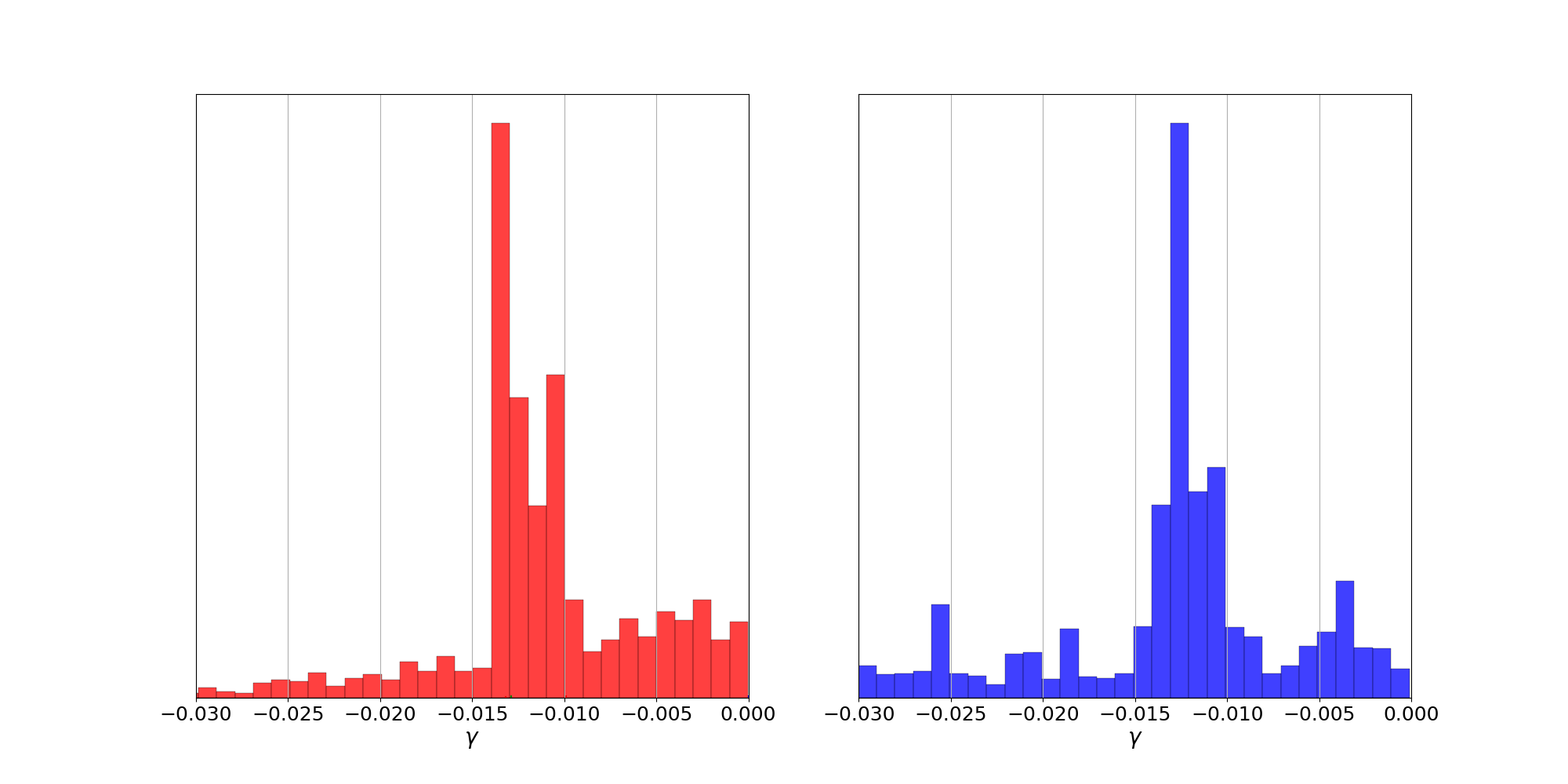}
    \caption{Comparison between the histograms of nested sampling and individuals through generations of the genetic algorithm for $\gamma$ parameter of Graduated Dark Energy model.}
    \label{fig:graduated_histograms}
\end{figure}

As mentioned above, some algorithms for Bayesian inference, such as multinest nested sampling, could explore the regions with these two maxima; however, most MCMC methods cannot achieve this task. Using genetic algorithms with the niching and sharing techniques, we can quickly find and explore the parameter space with these two optima without performing a Bayesian inference process; we can notice them in the histograms of  Figure \ref{fig:graduated_histograms}, in which the GA explore the regions of both modes of the $\gamma$ parameter; therefore we can have more confidence in the results of a genetic algorithm than a classical optimization method.

To conclude this section, it is worth noting that there are other multimodal cosmological models, mainly involving neutrinos and spatial curvature, documented 
in the literature \cite{kreisch2022atacama, camarena2023two, cedeno2021revisiting, park2019lambda, de2023current}, and worth exploring in future works where these techniques could prove valuable for conducting efficient and rapid assessments.

\begin{table}[t!]
    \captionsetup{justification=raggedright, singlelinecheck=false, font=scriptsize}
    \centering
    \scriptsize
    \begin{tabular}{|c|c|c|c|}
         \hline
         & Nested sampling & L-BFGS & Genetic \\
         \hline 
        $\Omega_m$ & 0.3264 &  0.2991  & 0.2959\\
        \hline 
         $h$ & 0.6947 & 0.6760  & 0.6765\\
         \hline 
         $\gamma$ & -0.0129  &  0.0000 & -0.0127 \\
         \hline 
         $-2\log\Like$ & 55.8700 &  60.5781 & 61.6997 \\
         \hline
    \end{tabular}
    \caption{Parameter estimation with nested sampling (sampling the posterior probability distribution function), L-BFGS, and genetic algorithm. In these cases, we only consider the maximum likelihood found in the three methods and their corresponding parameter values.}
    \label{tab:graduated}
\end{table}

\subsection{Derived functions}
\label{sec:cosmo_fgivenx}

As an additional application, taking advantage of the genetic algorithms nature, we can use the saved individuals along generations to maximize the likelihood function and calculate derived functions to analyze their phenomenological behavior. This technique is usually used with the samples of the posterior probability with Bayesian inference algorithms, mapping the sampling of an estimated parameter to another derived. For example, the library \texttt{fgivenx} \cite{handley2019fgivenx} allows this mapping. In the case of the individuals of a likelihood optimization using genetic algorithms, the statistical meaning of the plots is not directly related to the posterior probability function; however, it can provide an idea of the behavior of derived functions given the estimated parameters. 

\begin{figure}[t!]
    \captionsetup{justification=raggedright, singlelinecheck=false, font=footnotesize}
    \centering
    \includegraphics[trim= -2mm 0mm 12mm 0mm, clip, width=4.2cm, height=4cm]{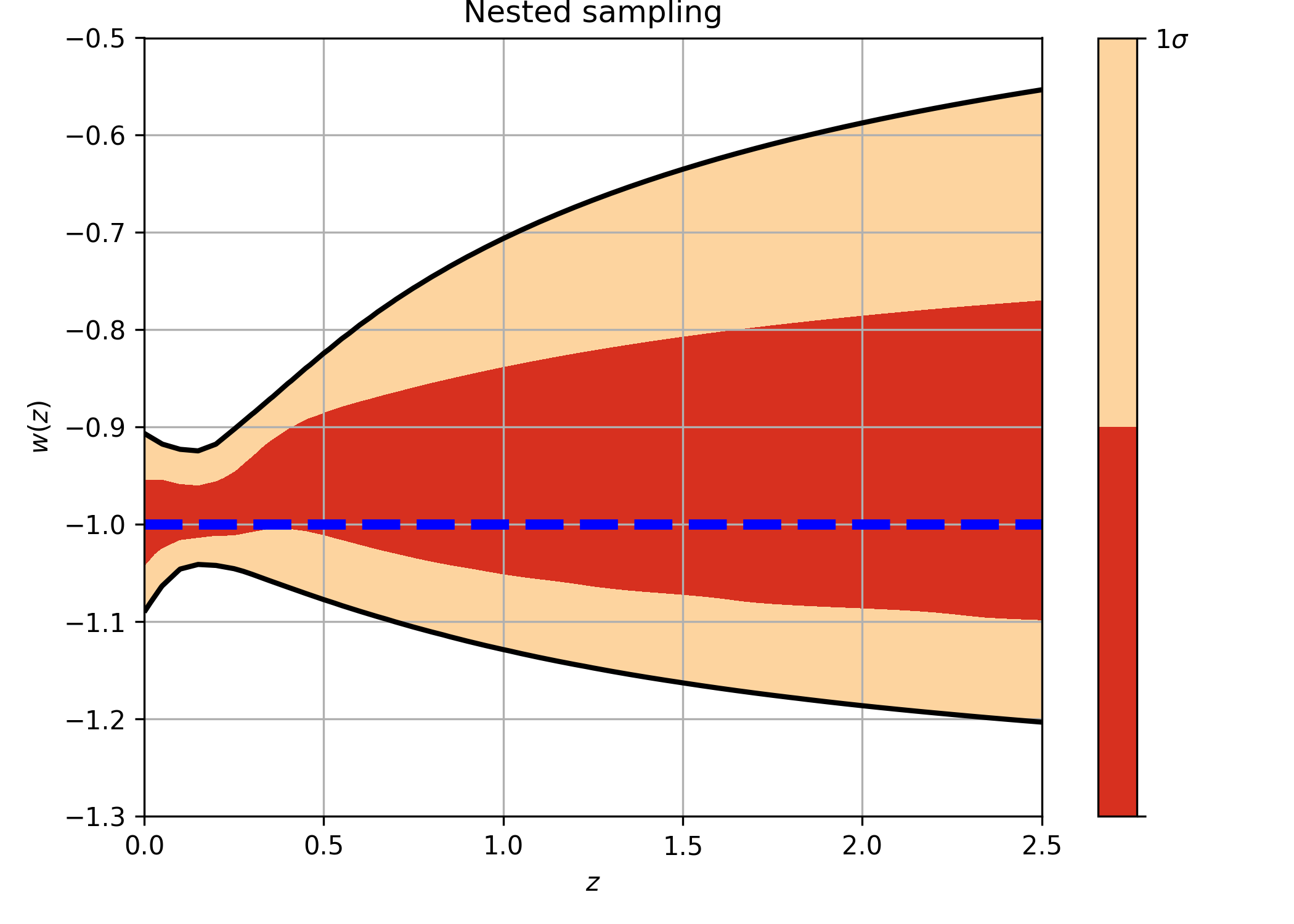}
    \includegraphics[trim= -2mm 0mm 12mm 0mm, clip, width=4.2cm, height=4cm]{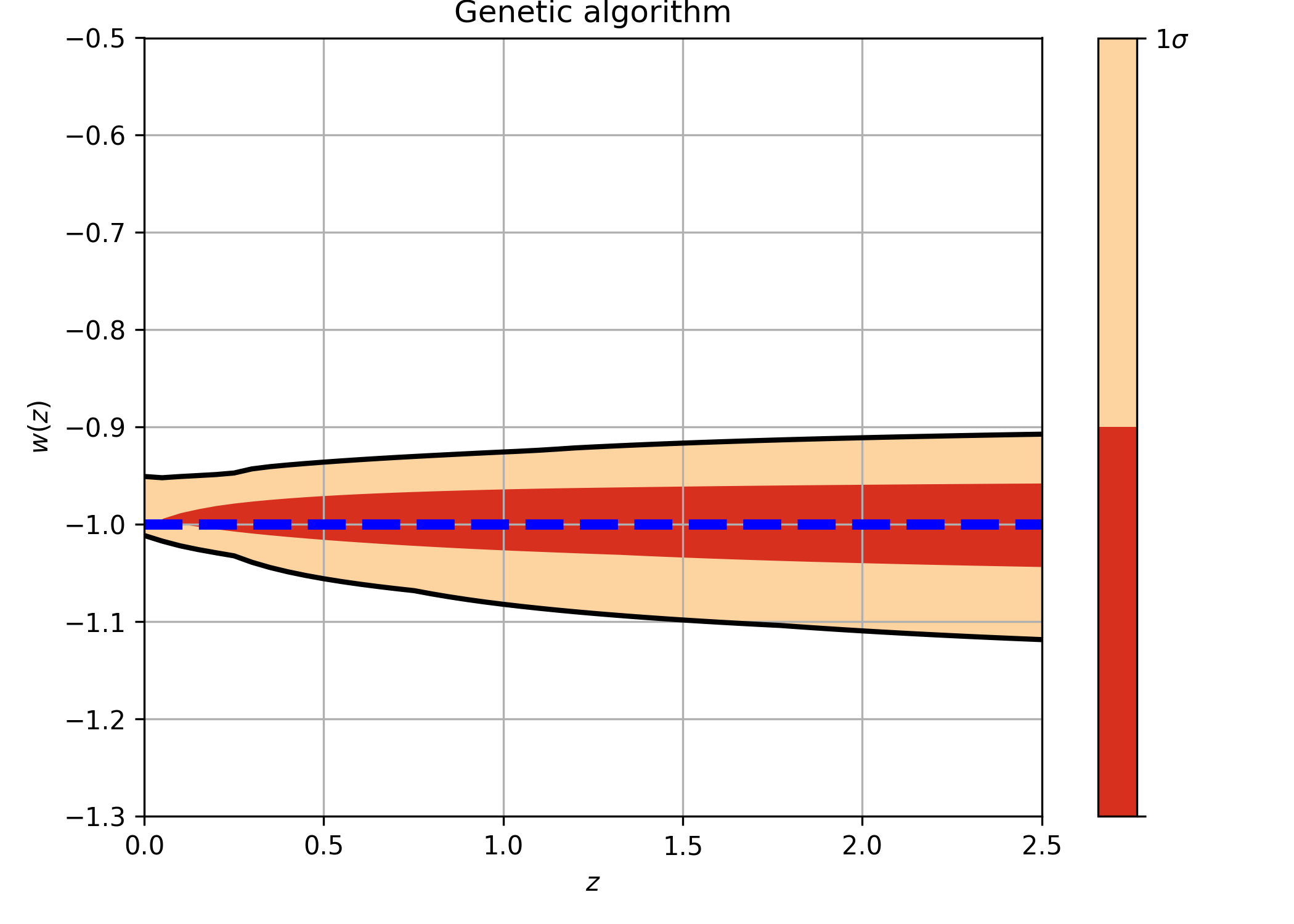}
    \caption{Equation of state for CPL model plotted with \texttt{fgivenx} from \textit{(left)} nested sampling, and \textit{(right)} genetic algorithms.}
    \label{fig:fgivenx}
\end{figure}

In Figure \ref{fig:fgivenx}, we compare the Equation of state reconstructed from the outputs of Section \ref{sec:cosmo_param_est} for the CPL model, we use the samples for the $w_0$ and $w_a$ from nested sampling, and the values of these same parameters from the historical of the individuals of the genetic algorithm population. We can notice that the behavior of the Equation of State, analyzing the darkest regions, is similar in both cases, and it suggests that for a quick test, we can use this technique with genetic algorithms. Regarding the confidence regions, because we are only optimizing the likelihood function with the genetic algorithms, we cannot have a formal way to estimate them correctly.

\section{\textbf{Conclusions}}
\label{sec:conclusiones}

In this study, we have leveraged genetic algorithms as an effective tool to estimate the free parameters of four cosmological models. Individuals generated in each genetic algorithm population have demonstrated the ability to achieve faster parameter estimates than those obtained using MCMC methods, thus reducing the number of likelihood function evaluations required. In addition, these genetic algorithms allow a rapid computation of derived parameters, which adds flexibility and efficiency to the estimation process.

However, it is important to note that genetic algorithms differ from Bayesian approaches in their sampling process. While MCMC methods fully sample the posterior probability function, genetic algorithms focus on maximizing the likelihood function. This distinction implies that genetic algorithms cannot directly provide confidence regions with the same statistical significance as Bayesian inference procedures. However, they offer significant advantages, such as faster speed and better results than other optimization methods, such as the L-BFGS algorithm.

Additionally, we have explored the usefulness of sharing and niche techniques in genetic algorithms, ensuring practical parameter space exploration, even in local or global optima. These features may be especially valuable in cosmology as a prior analysis to maximize the likelihood function before undertaking more computationally expensive Bayesian parameter estimation.

Throughout this paper, we can understand why genetic algorithms have been a very promising field of research over the last decades. Their flexibility allows their application in diverse tasks such as optimization, combinatorics, statistics, and even to speed up computational algorithms. \changes{The potential future applications of genetic algorithms in cosmological research are vast, with the presented study, we show the prospect of using them as a complement within cosmological data analysis. This is in agreement and complementary with existing research that also focuses on statistical applications of evolutionary computation \cite{surendran2022evolutionary, axiak2011evolution}; in our case, we have not proposed a novel method or algorithm, however, we have analyzed how to use GAs so that they can complement a traditional analysis of cosmological data and be an alternative to optimize the likelihood function. We are convinced that genetic algorithms are a great technique with diverse cosmological and statistical applications, for example, in a parallel work, we have explored their usefulness to improve cosmological neural reconstructions \cite{gomez2023neural} and to reduce the computational time of Bayesian inference routines \cite{Isidro}. Therefore, we are confident that genetic algorithms are an excellent complementary element to the cosmological data analysis toolkit.}

\section*{Acknowledgements}
J.A.V. acknowledges the support provided by FOSEC SEP-CONACYT Investigaci\'on B\'asica A1-S-21925, PRONACES-CONACYT/304001/2020, and~UNAM-DGAPA-PAPIIT IN117723. I.G.-V. appreciates the CONACYT postdoctoral grant and ICF-UNAM. RG-S acknowledges the support provided by SIP20230505-IPN, FORDECYT-PRONACES-CONACYT CF-MG-2558591, COFAA-IPN, and EDI-IPN grants.
 
\bibliographystyle{unsrt}
\bibliography{Bibliografia.bib}
\end{document}